\documentstyle[12pt,a4wide]{article}
\begin{document}
\author{B. Linet \thanks{E-mail: linet@celfi.phys.univ-tours.fr} \\
\small Laboratoire de Math\'ematiques et Physique Th\'eorique \\
\small CNRS/UPRES-A 6083, Universit\'e Fran\c{c}ois Rabelais \\
\small Facult\'e des Sciences et Techniques \\
\small Parc de Grandmont, 37200 TOURS, France}
\title{\bf Characteristic hypersurfaces \\
in a relativistic superfluid theory}
\date{}
\maketitle

\begin{abstract}
By discussing the Cauchy problem, we determine the covariant equation of 
the characteristic hypersurfaces in a relativistic superfluid theory.
\end{abstract}

\section{Introduction}

Phenomenological approaches to relativistic superfluidity had been initially
proposed by Rothen \cite{rot}, Dixon \cite{dix} and Israel \cite{isra,isrb}. 
Then, Khalatnikov and Lebedev \cite{khaa,khab} and Carter \cite{car1a,car1b} 
have developed general superfluid theories which are in fact equivalent as 
they have later shown \cite{car2a,car2b}. This is a relativistic 
generalization of the non-dissipative version of the Landau theory of 
superfluidity.

In the framework of this superfluid theory, Carter and Langlois \cite{car3} 
have studied the velocity of propagation of the sound following a method 
due to Hadamard. The story of infinitesimal disturbances, giving the sound 
speed, is described by an hypersurface for which the field variables are 
continuous but their derivatives are discontinuous  across this one. 
The problem is to find the equation of these characteristic hypersurfaces.
However, Carter and Langlois \cite{car3} do not solve the general case but
the low-temperature limit for a relativistic superfluid with
phonon-like excitation spectrum whose they have derived the Lagrangian.
Likewise, Vlasov \cite{vlaa,vlab} analyses the shock waves in this 
relativistic superfluid theory.

The purpose of the present work is to give the covariant equation
determining the characteristic hypersurfaces in this relativistic
superfluid theory. Our method is based on the fact that the Cauchy problem 
of the field equations has no unique solution
on a characteristic hypersurface. To do this, we consider the
covariant field equations of the superfluid in a spacetime with a
background metric $g_{\mu \nu}$ in a general coordinate system $(x^{\mu})$.
We attempt to solve the initial value problem on the hypersurface $x^1=0$
by finding the Cauchy data and by expressing merely the derivatives with
respect to $x^1$ in terms of the Cauchy data, method used in hydrodynamics 
for instance by Lichnerowicz \cite{lic}. If it is not possible to determine
uniquely the derivatives with respect to $x^1$ then $x^1=0$ is a 
characteristic hypersurface. Also, discontinuities of the normal derivatives 
of the field variables can occur across $x^1=0$. In another 
coordinate system $(y^{\mu})$, the equation $x^1=0$ will be $f(y^{\mu})=0$ 
and by this way we will find a covariant equation for the function $f$. 

The plan of the work is as follows. In section 2, we recall the field 
equations of the relativistic superfluid theory. We discuss the Cauchy problem 
in section 3. The determination of the
covariant equation of the characteristic hypersurfaces is
carried out in section 4. We add in section 5 some concluding remarks. 

\section{Field equations of the relativistic superfluid}

The relativistic superfluid theory of Carter is built on a particle
number $n^{\rho}$ and an entropy current $s^{\rho}$ which are conserved.
In a general coordinate system, these conservation laws can be written
\begin{equation}
\label{III}
\nabla_{\rho}n^{\rho}=0 \; ,
\end{equation}
\begin{equation}
\label{I}
\nabla_{\rho}s^{\rho}=0
\end{equation}
where $\nabla_{\sigma}$ denotes the covariante derivative associated with the 
metric $g_{\mu \nu}$. A master function $\Lambda$ of $n^{\rho}$ and
$s^{\rho}$ defines the chemical potentiel $\mu_{\sigma}$ and the
temperature $\theta_{\sigma}$ by the differential relation
$d\Lambda =\mu_{\rho}dn^{\rho}+\theta_{\rho}ds^{\rho}$.
The convective variational principle of Carter yields the equation
\begin{equation}
\label{IV}
s^{\sigma}(\nabla_{\sigma}\theta_{\rho}-\nabla_{\rho}\theta_{\sigma})=0
\end{equation}
and a similar equation for $\mu_{\sigma}$ but the assumption of 
superfluidity requires the compatible condition
\begin{equation}
\label{II}
\nabla_{\sigma}\mu_{\rho}-\nabla_{\rho}\mu_{\sigma}=0 \; .
\end{equation}

For our purpose, it is more convenient to introduce a Lagrangian $L$ having 
the expression 
$L=\Lambda -\mu_{\rho}n^{\rho}$ which is considered as a function of $s^{\rho}$
and $\mu_{\sigma}$ defining $\theta_{\sigma}$ and $n^{\rho}$ by the
differential relation $dL=\theta_{\rho}ds^{\rho}-n^{\rho}d\mu_{\rho}$.
The Lagrangian $L$ is in fact a function of the following scalars 
\begin{equation}
\label{scalar}
s^2=-s^{\rho}s_{\rho}\quad , \quad \mu^2=-\mu^{\rho}\mu_{\rho} \quad {\rm and}
\quad y^2=-s^{\rho}\mu_{\rho}
\end{equation}
defined for a signature $(-+++)$ of the metric.
In consequence, we can express $n^{\rho}$ and $\theta^{\rho}$ in terms
of $\mu_{\sigma}$ and $s^{\rho}$ in the following form
\begin{equation}
\label{i}
n^{\rho}=B\mu^{\rho}-As^{\rho} \quad {\rm and}\quad 
\theta^{\rho}=Cs^{\rho}+A\mu^{\rho}
\end{equation}
where the functions $A$, $B$ and $C$ of $s^2$, $\mu^2$ and $y^2$
are the partial derivatives of $L$ 
\begin{equation}
\label{d1}
B=2\frac{\partial L}{\partial \mu^2}\quad , \quad 
C=-2\frac{\partial L}{\partial s^2} \quad {\rm and}\quad
A=-\frac{\partial L}{\partial y^2}\; .
\end{equation}

We henceforth consider that the field variables of the relativistic superfluid
theory are $\mu_{\sigma}$ and $s^{\rho}$. From 
(\ref{III}) and (\ref{IV}) we derive the following equations
\begin{equation}
\label{1}
\mu^{\rho}\partial_{\rho}B+B\nabla_{\rho}\mu^{\rho}-s^{\rho}\partial_{\rho}A
=0 \; ,
\end{equation}
\begin{equation}
\label{2}
s^{\rho}s^{\sigma}\partial_{\sigma}C+Cs^{\sigma}\nabla_{\sigma}s^{\rho}+
\mu^{\rho}s^{\sigma}\partial_{\sigma}A+s^2g^{\rho \sigma}\partial_{\sigma}C+
\frac{1}{2}Cg^{\rho \sigma}\partial_{\sigma}s^2+y^2g^{\rho \sigma}
\partial_{\sigma}A=0
\end{equation}
where we have used (\ref{I}) and (\ref{II}).
The derivatives of $A$, $B$ and $C$ appearing in (\ref{1}) and (\ref{2}) have
the expressions 
\begin{equation}
\label{d2}
\partial_{\sigma}A=\frac{\partial A}{\partial s^2}\partial_{\sigma}s^2+
\frac{\partial A}{\partial \mu^2}\partial_{\sigma}\mu^2+
\frac{\partial A}{\partial y^2}\partial_{\sigma}y^2 \; .
\end{equation}
We notice the identities
\[
\frac{\partial A}{\partial \mu^2}=-\frac{1}{2}\frac{\partial B}{\partial y^2}
\quad , \quad \frac{\partial A}{\partial s^2}=\frac{1}{2}
\frac{\partial C}{\partial y^2} \quad {\rm and}\quad
\frac{\partial B}{\partial s^2}=-\frac{\partial C}{\partial \mu^2} \; .
\]

For a Lagrangian $L$ function of $s^2$, $\mu^2$ and $y^2$, the field 
equations are (\ref{I}), (\ref{II}), (\ref{1}) and (\ref{2}) with
definitions (\ref{d1}) and (\ref{d2}). 

\section{Cauchy problem}

We now discuss the Cauchy problem on the hypersurface $x^1=0$ for 
the field equations of the previous section, the field variables being 
$\mu_{\sigma}$ and $s^{\rho}$. The Cauchy data are $s^{\rho}$ and 
$\mu_{\sigma}$ on $x^1=0$. The problem is to determine 
$\partial_1\mu_{\sigma}$ and $\partial_1s^{\rho}$ on $x^1=0$ in terms of 
$s^{\rho}$ and $\mu_{\sigma}$ on $x^1=0$ and their derivatives with respect 
to $x^a$ $(a=0,2,3)$. Hereafter, we call $dC$ all quantities calculable in 
terms of the Cauchy data.

We immediately remark by virtue respectively of (\ref{I}) and (\ref{II}) that
\begin{equation}
\label{tri}
\partial_1\mu_a=dC \quad (a=0,2,3)\quad {\rm and} \quad \partial_1s^1=dC \; .
\end{equation}
According to (\ref{scalar}), we thus have
\begin{equation}
\label{E}
\partial_1\mu^2=-2\mu^1\partial_1\mu_1+dC \quad {\rm and}\quad
\partial_1s^2=-2s_a\partial_1s^a+dC \; .
\end{equation}
We insert (\ref{tri}) and (\ref{E}) in equations (\ref{1}) and (\ref{2})
for $\rho =a$ and we obtain four linear equations determining 
$\partial_1\mu_1$ and $\partial_1s^a$ in terms of the Cauchy data
\begin{eqnarray}
\label{matrice}
\left( \begin{array}{ll}
a & a_a \\
c_b & d_{ab} \\
\end{array} \right)
\left( \begin{array}{l}
\partial_1\mu_1 \\
\partial_1s^a \\
\end{array} \right) =
\left( \begin{array}{l}
dC \\
dC_b \\
\end{array} \right) \quad (a,b=0,2,3) \; .
\end{eqnarray}
The hypersurface $x^1=0$ is characteristic if the determinant of the matrix 
in (\ref{matrice}) vanishes. Unfortunately, it is not easy to find the
covariant equation of the characteristic hypersurfaces in this manner.

Instead of making use of (\ref{matrice}),
we are going to show that equations (\ref{1}) and (\ref{2}) determine 
$\partial_1\mu_1$, $\partial_1s^2$ and $\partial_1y^2$ 
in terms of the Cauchy data 
\begin{eqnarray}
\label{caract}
\left( \begin{array}{lll}
m_{11} & m_{12} & m_{13} \\
m_{21} & m_{22} & m_{23} \\
m_{31} & m_{32} & m_{33} \\
\end{array} \right)
\left( \begin{array}{l}
\partial_1\mu_1 \\
\partial_1s^2 \\
\partial_1y^2 \\
\end{array} \right) =
\left( \begin{array}{l}
dC \\
dC \\
dC \\
\end{array} \right) \; .
\end{eqnarray}
Furthermore, we will get
\begin{equation}
\label{prem}
s^1\partial_1s^a=dC
\end{equation}
when $\partial_1\mu_1$, $\partial_1s^2$ and $\partial_1y^2$ are known
from equations (\ref{caract}). 

We have always equation (\ref{1}) which gives
\begin{equation}
\label{A}
\mu^1\partial_1B+Bg^{11}\partial_1\mu_1-s^1\partial_1A=dC \; .
\end{equation}
We add equation (\ref{2}) for $\rho =1$ in the form
\begin{equation}
\label{B}
(s^1)^2\partial_1C+s^1\mu^1\partial_1A+s^2g^{11}\partial_1C+\frac{1}{2}Cg^{11}
\partial_1s^2+y^2g^{11}\partial_1A=dC \; .
\end{equation}
taking into account (\ref{tri}).
By contracting equation (\ref{2}) by $\mu_{\sigma}$, we obtain thereby
\begin{equation}
\label{S}
-y^2s^{\sigma}\partial_{\sigma}C-Cs^{\sigma}\partial_{\sigma}y^2-
Cs^{\sigma}s^{\rho}\nabla_{\sigma}\mu_{\rho}-\mu^2s^{\sigma}\partial_{\sigma}A
+s^2\mu^{\rho}\partial_{\rho}C+\frac{1}{2}C\mu^{\rho}\partial_{\rho}s^2+
y^2\mu^{\rho}\partial_{\rho}A=0
\end{equation}
where we have used the identity
\[
s^{\sigma}\mu_{\rho}\nabla_{\sigma}s^{\rho}=-s^{\sigma}\partial_{\sigma}y^2-
s^{\sigma}s^{\rho}\nabla_{\sigma}\mu_{\rho} \; .
\]
We now write down equation (\ref{S}) in the form
\begin{equation}
\label{C}
-y^2s^1\partial_1C-Cs^1\partial_1y^2-C(s^1)^2\partial_1\mu_1-\mu^2
s^1\partial_1A+s^2\mu^1\partial_1C+\frac{1}{2}C\mu^1\partial_1s^2+
y^2\mu^1\partial_1A=dC \; .
\end{equation}

In consequence, the three equations (\ref{A}), (\ref{B}) and (\ref{C}) give
a system for determining $\partial_1\mu_1$, $\partial_1s^2$ and 
$\partial_1y^2$ in terms of the Cauchy data . We obtain explicitly
the following system
\begin{eqnarray}
\label{E1}
\nonumber & & \left[ -2(\mu^1)^2\frac{\partial B}{\partial \mu^2}+2s^1\mu^1
\frac{\partial A}{\partial \mu^2}+Bg^{11}\right] \partial_1\mu_1 \\
& & +\left[ \mu^1\frac{\partial B}{\partial s^2}-s^1
\frac{\partial A}{\partial s^2}\right]
\partial_1s^2+ \left[ \mu^1\frac{\partial B}{\partial y^2}-s^1
\frac{\partial A}{\partial y^2}\right] \partial_1y^2=dC \; ,
\end{eqnarray}
\begin{eqnarray}
\label{E2}
\nonumber & & \left[ -2\mu^1(s^1)^2\frac{\partial C}{\partial \mu^2}-
2(\mu^1)^2s^1\frac{\partial A}{\partial \mu^2}-2\mu^1g^{11}
\frac{\partial C}{\partial \mu^2}s^2-2\mu^1g^{11}
\frac{\partial A}{\partial \mu^2}y^2 \right] \partial_1\mu_1 \\
\nonumber & & +\left[ (s^1)^2\frac{\partial C}{\partial s^2}+s^1\mu^1
\frac{\partial A}{\partial s^2}
+g^{11}\frac{\partial C}{\partial s^2}s^2
+g^{11}\frac{\partial A}{\partial s^2}y^2+\frac{1}{2}Cg^{11}\right] 
\partial_1s^2 \\
& & +\left[ (s^1)^2\frac{\partial C}{\partial y^2}+s^1\mu^1
\frac{\partial A}{\partial y^2}
+g^{11}\frac{\partial C}{\partial y^2}s^2+g^{11}\frac{\partial A}
{\partial y^2}y^2 \right] \partial_1y^2=dC \; ,
\end{eqnarray}
\begin{eqnarray}
\label{E3}
\nonumber & & \left[ 2\mu^1s^1\frac{\partial C}{\partial \mu^2}y^2
+2\mu^1s^1\frac{\partial A}{\partial \mu^2}\mu^2-2(\mu^1)^2
\frac{\partial C}{\partial \mu^2}s^2-2(\mu^1)^2
\frac{\partial A}{\partial \mu^2}y^2-C(s^1)^2 \right] \partial_1\mu_1 \\
\nonumber & & +\left[ -s^1\frac{\partial C}{\partial s^2}y^2-s^1
\frac{\partial A}{\partial s^2}\mu^2+\mu^1\frac{\partial C}{\partial s^2}s^2
+\mu^1\frac{\partial A}{\partial s^2}y^2+\frac{1}{2}C\mu^1 \right]
\partial_1s^2 \\
& & +\left[ -s^1\frac{\partial C}{\partial y^2}y^2-s^1
\frac{\partial A}{\partial y^2}\mu^2+\mu^1\frac{\partial C}{\partial y^2}s^2
+\mu^1\frac{\partial A}{\partial y^2}y^2
-Cs^1 \right] \partial_1y^2=dC
\end{eqnarray}
which corresponds to system (\ref{caract}).
Moreover, from equation (\ref{2}) for $\rho =a$ we find (\ref{prem}) when the
equations (\ref{E1}), (\ref{E2}) and (\ref{E3}) are satisfied.

\section{Covariant equation of the characteristic hypersurfaces}

When the condition $\det m_{ij}=0$ is verified it is not possible to determine
uniquely $\partial_1\mu_1$, $\partial_1s^1$ and $\partial_1y^2$ in terms of 
the Cauchy data. This condition defines the characteristic hypersurfaces.

In a coordinate system $(y^{\mu})$, for instance the Minkowskian coordinates,
the equation of an hypersurface is $f(y^{\mu})=0$. 
If we perform the change of coordinates
\begin{equation}
x^1=f(y^{\mu}) \quad {\rm and}\quad x^a=y^a \quad (a=0,2,3)
\end{equation}
then the equation $f(y^{\mu})=0$ becomes $x^1=0$. We can then apply the
results of the previous section in the coordinates $(x^{\mu})$.
It is crucial to notice that
\begin{equation}
s^1=s^{\alpha}\partial_{\alpha}f \quad , \quad \mu^1=
\mu^{\alpha}\partial_{\alpha}f \quad {\rm and}\quad 
g^{11}=g^{\alpha \beta}\partial_{\alpha}f\partial_{\beta}f
\end{equation}
where here $s^{\alpha}$, $\mu^{\alpha}$ and $g^{\alpha \beta}$ are the 
components in the coordinates $(y^{\mu})$. Furthermore, $s^2$, $\mu^2$ and
$y^2$ are scalars which can be written in the coordinates $(y^{\mu})$.

Now matrix $m_{ij}$ depends only on $s^1$, $\mu^1$, $g^{11}$ and the scalars
$s^2$, $\mu^2$ and $y^2$ taking into account the expressions of the 
functions $A$, $B$ and $C$, therefore the condition $\det m_{ij}=0$ gives 
a covariant equation for the function $f$
\begin{equation}
\label{relation}
{\cal F}(s^{\alpha}\partial_{\alpha}f,\mu^{\alpha}\partial_{\alpha}f,
g^{\alpha \beta}\partial_{\alpha}f\partial_{\beta}f,s^2,\mu^2,y^2)=0
\end{equation}
which is the desired 
equation of the characteristic hypersurfaces. According to (\ref{prem}) we 
must add the another possibility
\begin{equation}
s^{\alpha}\partial_{\alpha}f=0 \; .
\end{equation}
The speed of the sound $u$ with respect to the superfluid $\mu^{\rho}$ is 
given by
\[
u^2=\frac{(\mu^{\rho}\partial_{\rho}f)^2}{(\mu^2g^{\rho \sigma}+\mu^{\rho}
\mu^{\sigma})\partial_{\rho}f\partial_{\sigma}f}
\]
in units for which $c=1$.

\section{Conclusion}

Of course, it is difficult to exploit equation (\ref{relation})
determining the characteristic hypersurfaces. For obvious reasons of
causality, we must have $u<1$ and so
these characteristic hypersurfaces must be timelike, i.e.
\begin{equation}
\label{espace}
g^{\alpha \beta}\partial_{\alpha}f\partial_{\beta}f>0 \; .
\end{equation}
By solving (\ref{relation}), we must find  condition  (\ref{espace}) 
for all solutions to the field equations corresponding to a given Lagrangian
$L$. This restricts the generality of the function $L$.

We can apply our formalism in the case of the Lagrangian found by
Carter and Langlois \cite{car3} which has the expression
\begin{equation}
\label{etat}
L=L_0(\mu^2)-\frac{3k}{4}\left[ (c_{P}^{2}-1)\frac{y^4}{\mu^2}+s^2 
\right]^{2/3}
\end{equation}
where $k$ is a physical constant and $c_P$ the velocity of sound
for the fluid with one constituent defined by $L_0$ such that $c_P<1$. 
We assume that the 
relative velocity $v$ between $\mu^{\rho}$ and $s^{\rho}$ defined by
\[
1-v^2=\frac{s^2\mu^2}{y^4}
\]
is smaller than $c_P$. The temperature $T$ is given by
\[
T=k\frac{\mu}{y^2}\left[ (c_{P}^{2}-1)\frac{y^4}{\mu^2}+s^2 \right]^{2/3} \; .
\]
With expression (\ref{etat}), we can evaluate the functions
$A$, $B$ and $C$ and their derivatives that we express as
functions of $T$, $\mu$ and $v$. 
We take the low-temperature limit, $T\rightarrow 0$, keeping $\mu$ and $v$
fixed. Hence the condition $\det m_{ij}=0$ reduces to two separate conditions
\begin{equation}
\label{son1}
\lim_{T\rightarrow 0}m_{11}=0 \; ,
\end{equation}
\begin{equation}
\label{son2}
\lim_{T\rightarrow 0}T(m_{22}m_{33}-m_{23}m_{32})=0 \; .
\end{equation}
Condition (\ref{son1}) gives immediately $u_I=c_P$. 
Condition (\ref{son2}) is  complicated and we are not enabled to obtain a 
simple formula in order to compare with the one of Carter and Langlois
\cite{car3}. For the case in which $v=0$, we find $u_{II}=c_P/\sqrt{3}$
as expected.

\end{document}